# Timing Driven C-Slow Retiming on RTL for MultiCores on FPGAs

Tobias Strauch
*R&D*
*EDAptix*
*Adelgundenstr. 5, 80538 Munich, Germany*
email: tobias@edaptix.com

**Abstract.** In this paper C-Slow Retiming (CSR) on RTL is discussed. CSR multiplies the functionality of cores by adding the same number of registers into each path. The technique is ideal for FPGAs with their already existing registers. Previously publications are limited to adding registers on netlist level, which generates a lot of system verification problems and which is assumed to be the major drawback to use this technology in the modern multicore times. The paper shows how CSR can efficiently be done with timing driven automatic RTL modification. The methodology provided with this paper can be used as guidance for using CSR in high level synthesis (HLS). The paper shows the results of a CSR-ed complex RISC core on RTL implemented on FPGAs.

## Introduction

The increasing demands for higher performance and throughput of cores (CPUs, DSPs, peripherals, …) have led to various techniques. Pipelining a CPU engine (by using register insertion) is the most common example. Adding a number of pipelines into the instruction execution unit optimizes the throughput. Typically stall and flush signals are added to cope with the dependency of the instructions. Automatic pipelining of designs is outlined in [1].

All these methods are targeted to improve the performance of a single core. This paper demonstrates a method of how to use pipelining to **multiply** the functionality of a core. This is a fundamentally different outcome compared to what is known when designs are pipelined.

C-slow retiming uses pipelines to multiply the design behavior and to improve the latency of a core, as shown in [2]. This paper concentrates on CSR firstly as a method to increase performance per area when multiple equal cores are used (multicores) or if the application allows the usage of multiple equal cores (for instance, when multiple DSPs can increase the performance compared to one). CSR can also increase the throughput of a single design, especially for SoC interconnects, as shown in [3].

The key aspect of this paper is, that CSR is executed on RTL. The timing of the design is estimated and the RTL is modified for register insertion automatically. This has a serious impact on the flow and could be the key for a brighter acceptance of CSR.

CSR on RTL can then be applied in the IP development group and the CSR-ed core can be used by system architects, design developers and perhaps most importantly during the SoC implementation and verification process. The results of this paper can also be used as a guidance for CSR in HLS.

The "Theory of CSR" is explained in section 1. Section 2 shows the CSR on RTL methodology and section 3 shows the results of an CSR-ed complex RISC core in detail.

**1. Theory of CSR**

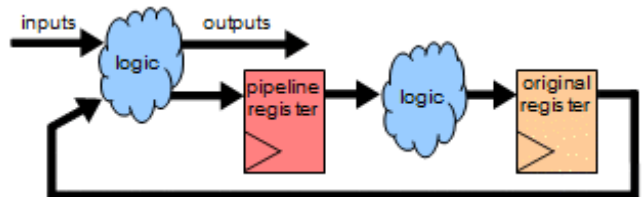

Figure 1. Two functional independent designs.

Figure 1 shows the basic structure of a sequential circuit. In this case, the original logic is sliced into two parts, and each original path has now 1 additional register. This results in 2 functional independent designs using the logic in a time sliced fashion. It shows how different parts of the logic are used during different cycles. The inputs and outputs are valid at the same time slice. The implemented register sets are called "system pipelines", SPs. Figure 1 shows one basic rule of CSR. There are only paths from the SPs to the original register set and from the register to the SPs.

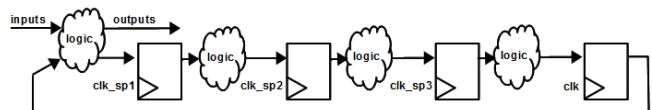

Figure 2. CSR core (CMF = 4) and individual clocks.

The number of SPs can be increased (Figure 2) and the resulting number of independent designs is identical to its multiplication factor, called "core multiplication factor", CMF. It is important to notice, that each SP gets an individual clock tree (clk_sp[n]). They can be used for power scaling techniques. If this feature is not needed, the clocks are connected on top level.

The main benefit is the multiplication of the core's functionality by only implementing registers instead of instantiating the core multiple times. The reduced **area** - compared to the area generated by individual cores - is a great advantage for ASICs and very attractive for FPGAs with their already existing registers. The CSR-ed design can run as many times faster as the number of the resulting segments (reduced by the additional setup and hold time of the SPs), but it also takes CMF times to execute the design. The **performance** remains the same.

It is questionable if CSR can be done efficiently on netlist level. CPUs have a high number of logic paths, which results in a long runtime of CSR for each synthesis P&R task, it needs to be specified, how the P&R-tool's functional modification of the core can be handled and how the input/output and memory access is served. CSR on RTL takes a few seconds and generates a defined RTL version of the core which can then be embedded in the system logic and the input/output/memory access can be optimized.

## 2. CSR on RTL Methodology

In this section the methodology of using CSR on RTL is presented. The novel approach is to elaborate the RTL source code and to estimate the timing of the elaborated design before synthesis. After the CSR optimization and register placement at the relevant signals in the design, the original RTL code is automatically modified. This modified RTL code can then be used for implementation, verification and synthesis for any given implementation, flow or technology.

The implemented registers are called "system pipelines", SPs. The number of independent cores after the SP insertion is CMF ("core multiplication factor").

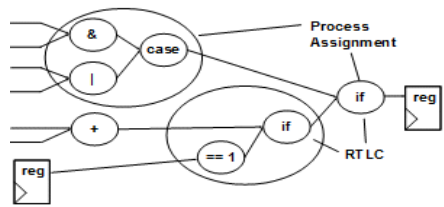

Figure 3. Registers, processes, assignments and RTLCs.

In the first step, the RTL code is read and elaborated. RTL is based on processes and assignments, which are built by RTL constructs (RTLC), as shown in Figure 3. The elaborated database holds the design with the predefined registers, processes, assignments, RTLCs and their connectivity.

Table 1. VHDL/Verilog Code Mapping to Logic Depth

| RTLC | VHDL | Verilog | 2-in Logic Depth |
|---:|---:|---:|---:|
| if | if the else | … ? … : … / if else | 1 |
| case | case (sel) when | case (sel) | log2\|sel\| |
| math | a + b, -, *, … | a + b, -, *, … | \|a\|, \|b\|, \|a+b\|, ... |
| comb | and, or, … | &, \|, … | 1 |
| unary | | &a, \|a, ^a | log2\|a\| |
| mux | a[i] | a[i] | log2\|a\|, log2\|i\| |
| demux | a[i] <= | a[i] <= | log2\|a\|, log2\|i\| |
| shift | Shl, shr | >>, << | \|a\|, \|i\| |

Table 1 shows the RTLC examples and their syntax in VHDL/Verilog. Each RTLC has a certain timing arc from the inputs to the output, which can be calculated based on the dimension of the RTLC (for instance, the dimension of the case select input). The resolution of the timing arc is "2-input logic depth" (2iLD). It is obvious, that the 2iLD can easily be calculated for each path through a process or assignment and for each register to register path. A bus is handles as a single path. For each RTLC, process or assignment, the worst case logic depth is taken. A 32x32-bit multiplier has therefore only 2 paths.

A simple algorithm now implements the SPs between processes and assignments. It thereby minimizes the 2iLD between SPs themselves as well as between SPs and registers. All SPs are placed at the outputs of the registers and then continuously moved through the logic, until an optimal distribution is reached.

Original Code:   assign lhs = rhs1 & rhs2;

Modified Code:   reg <rhs2_type> rhs2_sp1;
                 assign lhs = rhs1 & rhs2_sp1;
                 always @(posedge clk_sp1)
                 rhs2_sp1 <= #1 rsh2;

Figure 4. Code modification example.

The next step copies the original RTL files and modifies the RTL code to implement the SPs. Figure 4 gives an example of the code modifications. The left hand sided lhs signal depends on the two signals rhs1 and rhs2. If an SP should be placed at rhs2, the RTL line is modified and a register signal of the same type is implemented.

The final optional step considers previous STA results based on already placed and routed versions. It moves SPs to optimize the result (retiming). This step is skipped in the initial trial or if no back-annotated STA results should be considered.

**The optimal segmentation of the 2iLD paths leads to a reasonable good segmentation of the LUT/gate path.** The resulting CSR-ed design has very short paths (typically 3 to 4 LUTs for high CMFs). The short logic cones have a small set of permissible functions and a limited flexibility during technology mapping. Especially FPGAs with 6-input LUTs generate very predictable and reproducible results for small segments.

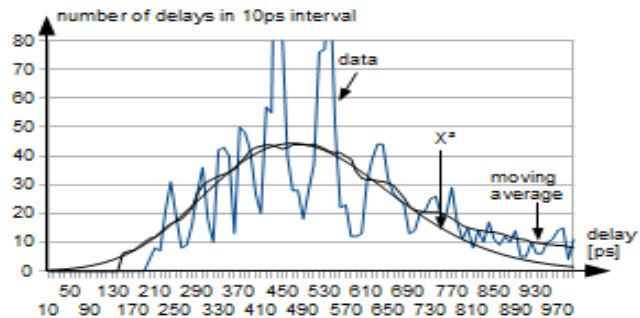

Figure 5. $X^2$ distribution for single LUT-net-pair delay.

ParaFPGA 2013, 10th - 13th September 2013, Munich, Germany

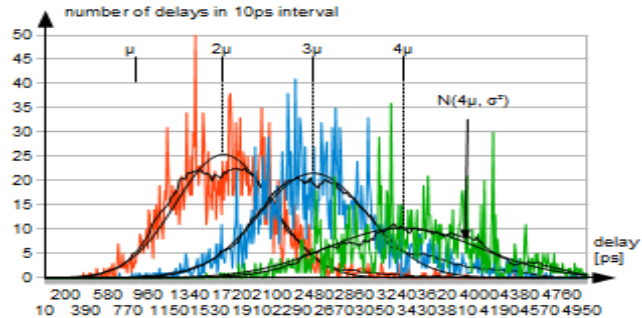

Figure 6. N(μ, σ²) for consecutive LUT-net-pairs

Empirical data of the example in section 3 now shows, that these small logic segments are very predictable in an unconstrained place and route step. A single LUT-net-pair delay usually follows a $X^2$ distribution (Figure 5), which leads with k>70 to a Gaussian distribution of consecutive LUT-net-pairs with $\mu_{LUT}$=825ps (Figure 6). Constrained physical synthesis now optimizes only very inefficient segments and is very limited due to the high number and very short logic cones.

So far it is not mentioned, how special function blocks (DSPs) are handled. An RTLC, which can be mapped to a special function block still gets a certain 2iLD. Empirical data shows, that timing arcs through a DSP-net-pair also follow a Gaussian with $\mu_{DSP}$=1520ps (Figure 6). This relation $\mu_{LUT}/\mu_{DSP}$ is reflected in the calculation of the 2iLD of special function blocks in FPGAs.

Memory in-, out- and throughput timing is also calculated based on 2iLD. For high CMFs, memories have a higher chance to be on the critical path and they typically dominate the critical CSR-ed path as well (see example in section 3). They are then replaced with pipelined memories.

## 3. RISC Processor Example

This section describes the CSR on RTL of a RISC core. The original code is taken from the OpenCores' OR1200 project [4]. The OR1200 is a 32-bit scalar RISC with Harvard micro-architecture, 5 stage integer pipeline, virtual memory support (MMU), basic DSP capabilities, TLB, instruction and data cache. The results are uploaded and accessible at [4].

*3.1. Slice Utilization and Performance for FPGAs*

The next Tables show the area and timing results for a Virtex5 device from Xilinx. In general, ISE 11.1 with the place and route effort option "standard" is used. The results are based on a Virtex5 device (xc5vlx50-1ff676, package FF676, speed grade -1). One implemented OR1200 core reaches 13.853ns (72.2MHz) on this device. The "Achieved Timing" number is the "Data Path Delay" number taken from the first report of the timing report (.twr). The "Theoretical Timing" considers the additional delays introduced by the SPs. The sum of setup time (Tas) and the hold time (Tcko) is 0.400ns. The theoretical achievable timing is shown in Figure 7.

$$\text{CMF} == 2: (13.853\text{ns} + 0.400\text{ns})/2 = 7.126\text{ns} (193\% \text{ of } 13.853\text{ns})$$
$$\text{CMF} == 3: (13.853\text{ns} + 0.800\text{ns})/3 = 4.884\text{ns} (282\% \text{ of } 13.853\text{ns})$$
$$\text{CMF} == 4: (13.853\text{ns} + 1.200\text{ns})/4 = 3.763\text{ns} (367\% \text{ of } 13.853\text{ns})$$

Figure 7. Theoretical achievable timing.

**Table 2.** Utilization of Virtex5 Device

| CMF | FF | Slice LUTs | Occupied Slices |
|---|---|---|---|
| 1 (orig) | 1239 | 3663 (12%) | 1131 (15%) |
| 2 | 3048 | 4535 (15%) | 1414 (19%) |
| 3 | 4153 | 5602 (19%) | 1594 (22%) |
| 4 | 4777 | 6286 (21%) | 1773 (24%) |

Table 2 shows the utilization results of the implemented CSR-ed OR1200 core. It lists how the numbers of registers (FF), slice look up tables (LUTs) and occupied slices increase with rising CMF. One core occupies 15% of the FPGA slices, whereas a CSR-ed core with CMF = 4 (4 independent cores) only occupies 24% of the FPGA slices. This is less than twice the original size, which is a significant area improvement.

**Table 3.** Timing of Virtex5 Device

| CMF | Theoretical Timing | Constraint | Achieved Timing | LUT Levels |
|---|---|---|---|---|
| 1 (orig) | n/a | 13.5ns | 13.853ns | 12 |
| 2 | 7.126ns | 7.120ns | 7.327ns | 7 |
| 3 | 4.884ns | 4.880ns | 5.398ns | R-Out + 3 |
| 4 | 3.763ns | 3.763ns | 4.902ns | R-Out + 1 |

**Table 4.** Relative Utilization and Performance

| CMF | Slice LUTs | Relative Utilization | Relative Performance | Timing Ratio | PpS [kHz] |
|---|---|---|---|---|---|
| 1 (orig) | 1 | 1 | 1 | 1 | 63.8 |
| 2 | 1.23 | 1.25 | 1.88 | 0.97 | 96.1 |
| 3 | 1.52 | 1.40 | 2.56 | 0.90 | 116 |
| 4 | 1.71 | 1.56 | 2.82 | 0.76 | 114 |

Table 3 shows the theoretical timing (Figure 7), the constraint used for syntheses and the achieved timing with the resulting LUT levels of the critical path. R-Out in Table 3 indicates a RAM output. The number of used DSP48Es for the 32x32-bit multiplier remains 4.

Table 4 can be read as follows. With CMF = 2, the number of slice LUTs rises by 23% and the number of occupied slices by 25%. The performance increases by 88%, which is 97% of the theoretical achievable timing. The performance per slice (PpS) is 96.1kHz. In non CSR-ed designs the area multiplies according to the CMF and the PpS index remains constant.

The average delay of a Virtex5 LUT-net-pair delay sum is assumed to be $\mu_{LUT}$=825ps. The difference between theoretical achievable timing and achieved timing is within this granularity of one LUT-net-pair. It can be argued, that the increased area of up to 56% should also be considered (CWLM). This worsens the achievable theoretical timing and reduces the optimization gap. The difference between the timing with and without optimizations based on back-annotated data is also within the granularity of one LUT-net-pair. This shows the efficiency of the proposed CSR on RTL method. CSR can be executed once and the RTL is available as functional correct IP for the implementation on system level. **Further timing optimizations are then done based on a design specific implementation with standard retiming algorithms.**

The device is relatively underutilized (15%). Other examples with higher utilization show, that the CSR-ed core can be better packed (less increase of occupied slices for the CSR-ed core). The following results are based on an AVR core [4] implemented on a Spartan3 device (XS3S200a, package FG320, speed grade -4). This is the smallest device for a single AVR core (occupied slices 59%, used 4-input-LUTs 48%) and it is not possible to implement 2 or more individual AVR cores on this device, because the number of slices are with 2012 out of 1792 available slices overmapped (112%).

**Table 5.** Utilization ans Timing of Spartan3 Device

| CMF | FF | 4-input LUTs | Occupied Slices | Achieved Timing | PpS [kHz] |
|---|---|---|---|---|---|
| 1 (orig) | 463 | 1748 (48%) | 1062 (59%) | 24.914ns | 37.7 |
| 2 | 1125 | 2344 (65%) | 1512 (84%) | 14.763ns | 44.8 |
| 3 | 1603 | 2691 (75%) | 1790 (99%) | 12.400ns | 45.0 |
| 4 | 1716 | 2990 (83%) | 1790 (99%) | 11.290ns | 49.4 |

Table 5 shows the number of FFs, 4-inputs LUTs and occupied slices of the CSR-ed AVR core. Although it is not possible to implement 2 single AVR cores on this device, the CSR on RTL method enables the implementation of a CSR-ed AVR core with CMF = 4, which means that 4 independent AVR core behaviors are available. The device has already an utilization of 99% for CMF = 3, but still registers can be implemented to reach a CMF of 4. The achieved timing and the PpS improved with all CMFs.

*3.2. Area Ratio for ASICs*

If the CSR-ed OR1200 core is implemented on an ASIC, the size of the combinatorial logic (gates) remains almost the same, only the number of registers increases. This number should not be simply multiplied, because the registers of the new implemented SPs are located at internal signals. An m+n-adder adds m+n registers, if the registers are placed at the inputs, but only max(m, n)+1, if they are placed at the outputs of the adder logic.

**Table 6.** Area Ratio for ASICs

| CMF | FF | Relativ FF | Relative Area with 42(gate)/58(FF) Ratio |
|---|---|---|---|
| 1 | 1239 | 1 | 1 |
| 2 | 2995 | 2.42 | 1.59 |
| 3 | 3769 | 3.04 | 1.85 |
| 4 | 4244 | 3.43 | 2.01 |

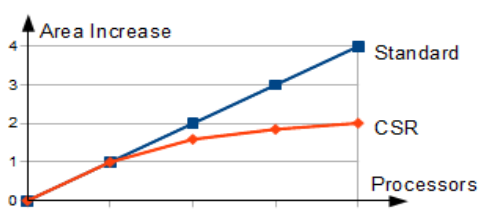
Figure 8. Area increase for multicores using CSR.

Table 6 shows the number of registers implemented in the OR1200 core without the FPGA specific timing optimizations. Figure 8 shows the results of Table 6 graphically. The number of registers increases by 142%, 204% and 243%. If the ratio of register area and combinatorial logic is set to 42/58 (42% register area and 58% combinatorial logic), which is based on a synthesis report using the lsi_10k library, the area increases by 59%, 85% or 101% of the original area. In other words, the area of the logic (excluding memories) of 4 CSR-ed OR1200 cores is only half as large as the one of 4 individual OR1200 implementations. This is a significant area reduction.

Area optimized shift registers and/or non-scan-FF can be used. CSR based multiprocessor systems have less gate count and smaller logic cones and therefore reduce test costs dramatically.

### 4. Conclusion

In this paper we presented the novel approach to use CSR on RTL. It is shown how CSR-ed cores generate area reduced multicores. CSR clearly changes the behavior of a core and can only be fully utilized, if the CSR-ed core is embedded in a new logic environment. *With the right wrapper logic, the CSR-ed core then behaves exactly as the original core, but multiple and functional independent versions are available.* This has a dramatic impact on the flow, which makes it mandatory to have a solution on higher level such as RTL. The CSR-ed version can be used as a new core in the design and verification process.

The authors doubt, that these modification can be done efficiently on netlists at all (the OR1200 has $8*10^7$ paths) and expect major flow problems in pure netlist based approaches. The P&R-tool should not modify the core's behavior and should not need to do the modifications over and over again. The runtime of the tool used in this paper is within a few seconds.

An algorithm for CSR on RTL is shown and utilization, area and timing results of a comprehensive RISC processor are given. It shows that the CSR-ed core can provide 4 times the behavior for the area and utilization costs of 2 individual implementations. The timing for the CSR-ed RISC example on RTL reaches the theoretical achievable timing within the granularity of one LUT, eliminating the need to do the modifications on netlist level. The authors are not aware of any results to compare with. Multicores usually multiply the area as well (Figure 8) and have a constant performance per area (slice) ratio (Table 4). The methodology presented in this paper can also be used if CSR is used in HLS.

The examples are based on processors. Although CSR is not limited to processors (the complete SoC-bus systems as well as peripherals and DSPs can be CSR-ed as well), it can easily be imagined, how the examples can be embedded in a multithreading processor design.

**REFERENCES**
[1] Kroening D. and Paul W. 2001. Automated Pipeline Design. Proceedings of the Design Automation Conference 2001, June 18-22, Las Vegas, Nevada, USA, pages 810-815
[2] Weaver N., Markovskiy Y., Patel Y., and Wawrzynek J. 2003. Post-Placement C-slow Retiming for the Xilinx Virtex FPGA. Proceedings of the 11th intl. Symposium on FPGAs 2003, Feb. 23-25, 2003, Monterey, USA, pages 185-194.
[3] Hong J., Fan Y., and Zhang Z. 2004. Architecture-Level Synthesis for Automatic Interconnect Pipelining. Proceedings of the Design Automation Conference 2004, June 7-11, San Diego, CA, USA, pages 602-607
[4] 2007, Stockholm, Sweden, ORSoC AB, Projects. Available: www.opencores.org/projects